\documentclass[aps,prl,twocolumn,,showpacs,superscriptaddress]{revtex4-1}
\usepackage{graphicx}
\usepackage[%
  pdftitle={Cooper Pair Breakup in YBCO under Strong Terahertz Fields},%
  pdfauthor={Andreas Glossner, Caihong Zhang, Shinya Kikuta, Iwao Kawayama, Hironaru Murakami, Paul Müller and Masayoshi Tonouchi},%
  pdfsubject={Cooper Pair Breakup in YBCO under Strong Terahertz Fields},%
]{hyperref}

\begin{document}
\hyphenation{super-conductor}
\hyphenation{super-conducting}
\hyphenation{tera-hertz}
\hyphenation{future}
\hyphenation{Osaka}
\hyphenation{panels}
%\preprint{}

\title{Cooper Pair Breakup in YBa$_2$Cu$_3$O$_{7-\delta}$ under Strong Terahertz Fields}


\author{A. Glossner}
\affiliation{Institute of Laser Engineering, Osaka University, Yamada-oka 26, Suita, Osaka 565-0871, Japan} 
\affiliation{Department of Physics, Universit\"{a}t Erlangen-N\"{u}rnberg, Erwin-Rommel-Stra{\ss}e 1, 91058 Erlangen, Germany}

\author{C. Zhang}
\author{S. Kikuta}
\author{I. Kawayama}
\author{H. Murakami}
\affiliation{Institute of Laser Engineering, Osaka University, Yamada-oka 26, Suita, Osaka 565-0871, Japan}

\author{P. M\"{u}ller}
\affiliation{Department of Physics, Universit\"{a}t Erlangen-N\"{u}rnberg, Erwin-Rommel-Stra{\ss}e 1, 91058 Erlangen, Germany}

\author{M. Tonouchi}
\affiliation{Institute of Laser Engineering, Osaka University, Yamada-oka 26, Suita, Osaka 565-0871, Japan}
%\thanks{}


\date{\today}

\begin{abstract}
We show that strong electric fields of $\sim30\,$kV\,cm$^{-1}$ at terahertz frequencies can significantly weaken the superconducting characteristics of cuprate superconductors. High-power terahertz time-domain spectroscopy (THz-TDS) was used to investigate the in-plane conductivity of YBa$_2$Cu$_3$O$_{7-\delta}$ (YBCO) with highly intense single-cycle terahertz pulses. Even though the terahertz photon energy ($\sim 1.5\,$meV) was significantly smaller than the energy gap in YBCO ($\sim 20$-$30\,$meV), the optical conductivity was highly sensitive to the field strength of the applied terahertz transients. Possibly, this is due to an ultrafast, field-induced modification of the superconductor's effective coupling function, leading to a massive Cooper pair breakup. The effect was evident for several YBCO thin films on MgO and LSAT substrates.
\end{abstract}

%\keywords{}

\maketitle



Conventional terahertz spectroscopy permits a direct measurement of the temperature and frequency dependence of the conductivity and has allowed for a detailed examination of parameters such as the plasma frequency, the superfluid density or the London penetration depth~\cite{1,2,3,4,5,6,7,8,9,10,11}. Optical pump-terahertz probe studies have established a better understanding of the nonequilibrium behavior of cuprate superconductors, particularly the superconducting pair and quasiparticle relaxation dynamics after femtosecond excitation~\cite{4, 5, 7}. Recently, Pashkin~\emph{et al.}~\cite{12} have investigated the electron-lattice interactions in YBCO after mid-infrared photoexcitation on a sub-picosecond time scale. They concluded that the pump energy leads to a competition between the population of phonons and the depletion of the superconducting condensate~\cite{12, 13}. Fausti~\emph{et al.} have used the pump-probe technique to induce superconductivity in non-superconducting La$_{1.675}$Eu$_{0.2}$Sr$_{0.125}$CuO~\cite{14}.

We have employed high-power terahertz time-domain spectroscopy to investigate the time-resolved behavior of the high-$T_c$ superconductor YBCO in the presence of strong terahertz electric fields. With this method, we were able to separate the field-induced changes from other effects, particularly heating and photon-induced Cooper pair breaking. Pulses at ps-duration only require energies in the range of 0.1\,$\mu$J to reach high peak electric field strengths, i.e. heating is avoided. Similarly, the spectral range of the pulses between 0.2 and 1\,THz with a center frequency of 0.35\,THz ruled out massive photon-induced pair breaking. In combination with the direct accessibility of the conductivity, high-power terahertz time-domain spectroscopy (THz-TDS) therefore provided a powerful tool to investigate an aspect of high-temperature superconductivity that has been elusive so far: the behavior of YBCO when exposed to terahertz-frequency transients with ultra-strong peak electric fields.\looseness=1

\begin{figure}
  \includegraphics[width=0.9\columnwidth]{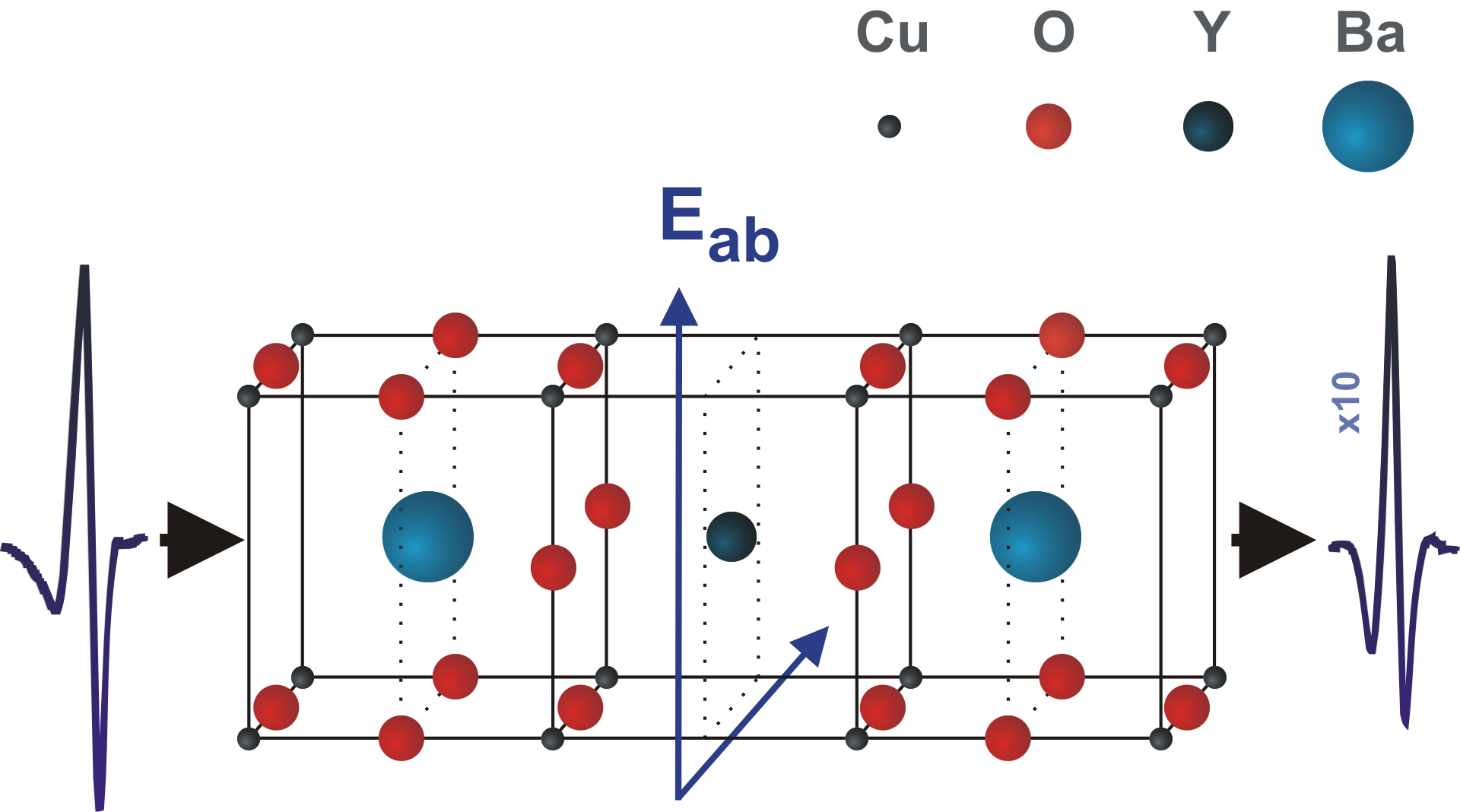}
 \caption{Transmission of intense, single-cycle THz pulses through YBa$_2$Cu$_3$O$_{7-\delta}$. Thin films were probed with THz pulses of up to $30\,$kV\,cm$^{-1}$ field strength. The THz waves propagate in $c$-direction, i.e. the electric field was polarized parallel to the $a$-$b$ planes of the superconducting film.}\label{fig1}
\end{figure}

By means of the tilted-pulse-front method~\cite{15, 16}, we generated highly intense, single-cycle terahertz pulses through optical rectification in a LiNbO$_3$ crystal which were then transmitted through the YBCO thin films (Fig.~\ref{fig1}). Thereby, it was possible to probe the films with peak terahertz electric fields of approximately $30\,$kV\,cm$^{-1}$. Our experiment is a consequence of the rapid progress in the generation of intense terahertz radiation in the last few years, which has established the field of nonlinear terahertz science~\cite{2, 15,16,17}. The polarization was parallel to the $a$-$b$ planes. A high signal-to-noise ratio permitted the investigation of the field-dependent behavior of YBCO at low temperatures, at which the transmission is small due to the high surface impedance of the superconducting thin films. The measurements were conducted on optimally doped, expitaxially-grown YBCO thin film samples of 45\,nm and 120\,nm thickness. The substrates were 500\,$\mu$m (LaAlO$_3$)$_{0.3}$\,(Sr$_2$AlTaO$_6$)$_{0.7}$ (LSAT) and MgO single crystals, respectively, and the critical temperature of the thin films was 88\,K.

\begin{figure}
  \includegraphics[width=0.94\columnwidth]{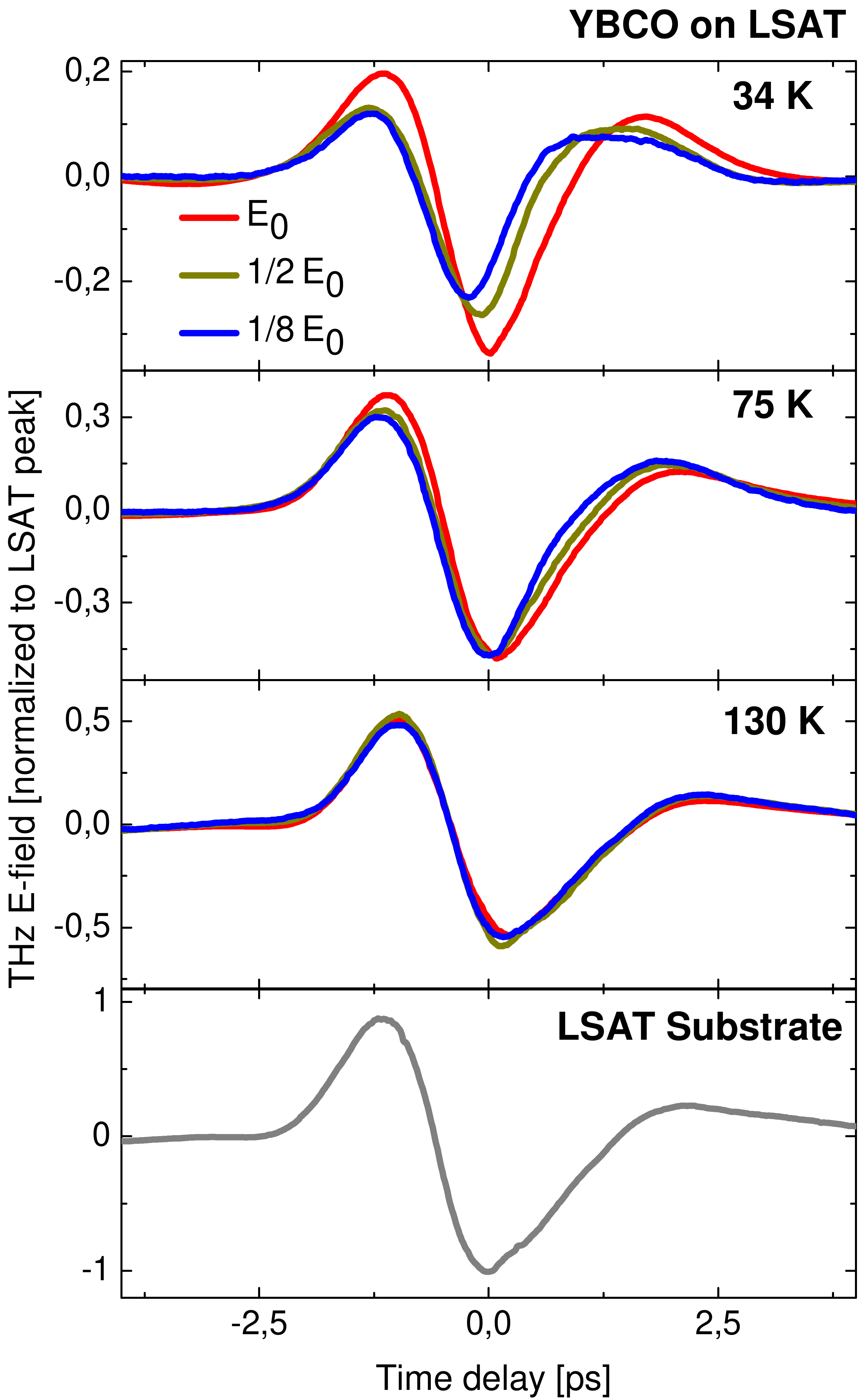}
 \caption{Terahertz pulses transmitted through a 45\,nm YBa$_2$Cu$_3$O$_{7-\delta}$ film on LSAT. The time-domain transients, measured by electro-optic sampling, are shown for temperatures well below (upper two panels) and above the superconducting transition temperature. E$_0$ corresponds to a field strength of approximately $30\,$kV\,cm$^{-1}$. All curves are normalized to the peaks of the bare substrate measurements (bottom panel). As the field strength is increased at 34\,K and 75\,K, a field-induced modification of the effective coupling function leads to a massive Cooper pair break-up. The transmission increases and the shape of the curve becomes more similar to the non-superconducting case (red curves in upper two panels). At temperatures below $T_c$, the superconducting characteristics therefore sensitively depend on the field strength of the applied terahertz transients. The effect could also be reproduced on a 120\,nm YBCO thin film on MgO substrate.}\label{fig2}
\end{figure}

A first feature of the field-dependent behavior of the YBCO thin films is already given by the bare time-domain information obtained from the THz-TDS measurements. Figure~\ref{fig2} shows the time-resolved terahertz signal after transmission of the pulse through a YBCO thin film on LSAT for temperatures below and above the critical temperature, respectively. At 130\,K in the non-superconducting state, the response is hardly sensitive to the field strength of the probing terahertz pulses. The wave shape is almost identical to the one in the bare-substrate case. Similarly, the low-field, low-temperature response (blue curve in upper panel, Fig.~\ref{fig2}) is in agreement with previous time-resolved terahertz studies~\cite{6, 18}. In this state, the transmission is significantly reduced. The superconductor's kinetic inductance leads to a clear modification of the waveform with respect to the reference pulse~\cite{6}. This behavior changes, however, when applying the full terahertz field of roughly $30\,$kV\,cm$^{-1}$ at 34\,K and 75\,K. The transmission increases strongly with increasing field strength (red curve in Fig.~\ref{fig2}). The waveform becomes similar to the response in the non-superconducting state. This suggests a decrease of the kinetic inductance and a clear damping of the superconductor's response in strong terahertz electric fields.

\begin{figure}
  \includegraphics[width=0.9\columnwidth]{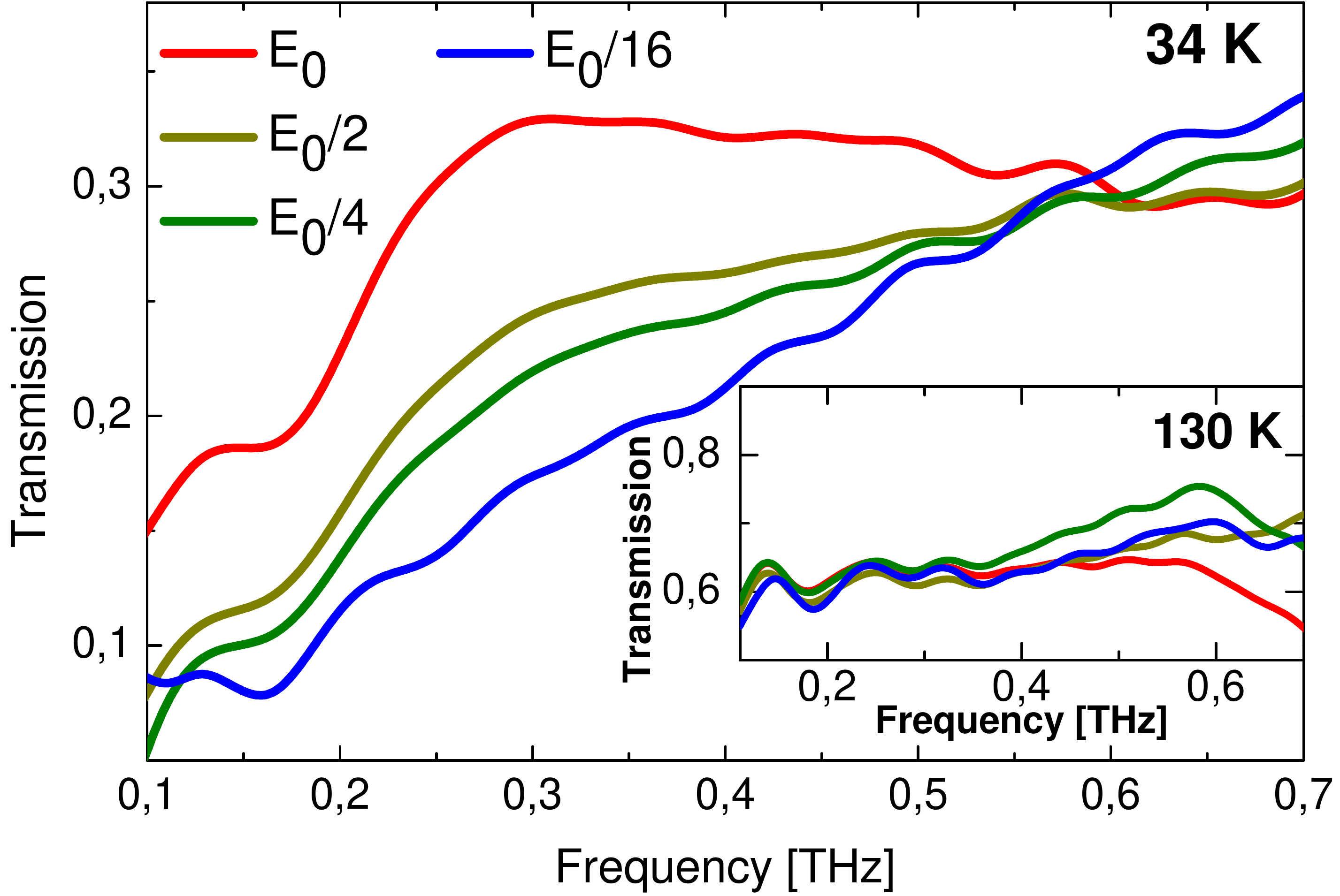}
 \caption{Transmission at different field strengths, measured at 34\,K and 130\,K (inset). The frequency dependence of the transmission, normalized to the bare substrate, is displayed for different probe field strengths. At temperatures below $T_c$, the damping of the superconductivity under strong fields causes the film to become increasingly transparent at low frequencies (main panel, red curve).}\label{fig3}
\end{figure}

\begin{figure*}
  \includegraphics[width=1.0\textwidth]{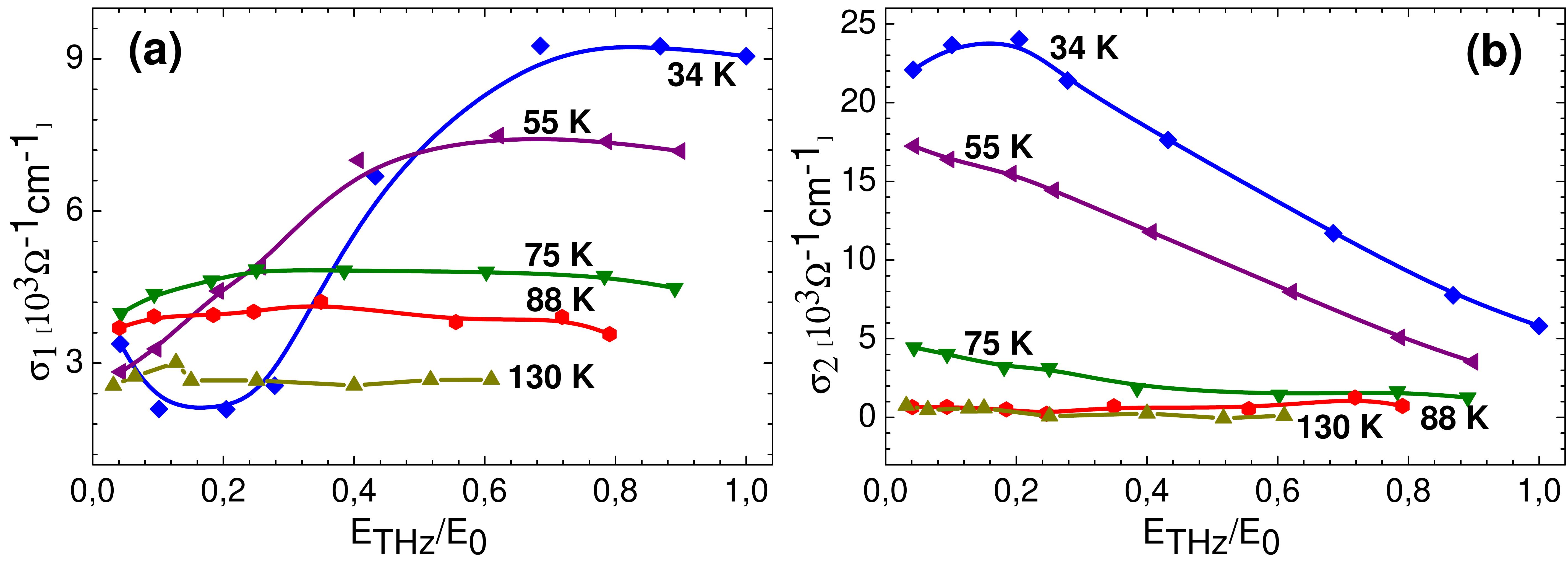}
 \caption{Field-dependent optical conductivity. For temperatures below and above $T_c$, the complex conductivity is shown as a function of the THz field strength. The values at 0.4\,THz are displayed. At temperatures below $T_c$, note the strong increase of $\sigma_1$ at high field strengths. This is accompanied by a strong drop of $\sigma_2$, indicative of a massive field-induced Cooper pair breakup. For the critical temperature of 88\,K (red hexagons) and for 130\,K (yellow triangles), the complex conductivity is almost terahertz field-independent. This behavior could be observed for the entire frequency range and for all samples. Lines serve to guide the eye.}\label{fig4}
\end{figure*}

A look at the frequency-dependence of the transmission at 34\,K, shown in Fig.~\ref{fig3}, provides further insight into this behavior. As expected from the time-domain data, for frequencies in the range between 0.1 and 0.6\,THz, the transmission increases with higher terahertz field strengths. We note that the shift is accompanied by a change in the frequency dependence. For weak electric fields, the transmission increases linearly with frequency, in accordance with previous findings with low-power THz-TDS~\cite{6}. The kinetic inductance of the superconducting charge carriers causes the thin film to act like a high pass filter, i.e. higher frequencies experience less attenuation~\cite{6, 18}. In the presence of strong terahertz fields, however, the transmission increases at lower frequencies. At 0.35\,THz, it exhibits a flattening close to the center frequency of the incident pulses while the transmission at higher frequencies remains almost unchanged (Fig.~\ref{fig3}). Thus, the high-pass filter model does not apply for high terahertz field strengths, which suggests a strong reduction of the kinetic inductance.

It is evident that this modification of the kinetic inductance is caused by a field-induced breakup of Cooper pairs. Figure~\ref{fig4} shows the field-dependence of the complex conductivity for different temperatures. For increasing terahertz fields, there is a clear drop of $\sigma_2$ and an increase of $\sigma_1$ at temperatures below $T_c$. This behavior was observed in all the samples as well as in the entire frequency range. This is remarkable because it indicates a massive Cooper pair breakup, in spite of the fact that the photon energy was well below the energy gap of YBCO. A peak frequency of 0.35\,THz corresponds to a photon energy of approximately 1.5\,meV, while the energy gap of YBCO is in the range of 20\,meV~\cite{19}. Similarly, the pulse energies are too low for bare heating. This requires an explanation different from conventional photon-induced or thermally excited Cooper pair breakup.

The terahertz transients with peak electric field strengths of roughly $30\,$kV\,cm$^{-1}$ and the evident field dependence of the superconducting properties suggest a field-induced modification of the superconductor's effective coupling function. As a result, a Cooper pair breakup is taking place even well below the energy gap of the superconductor. The scale of this effect is determined by the strength of the applied terahertz field.

\begin{figure}
  \includegraphics[width=1.0\columnwidth]{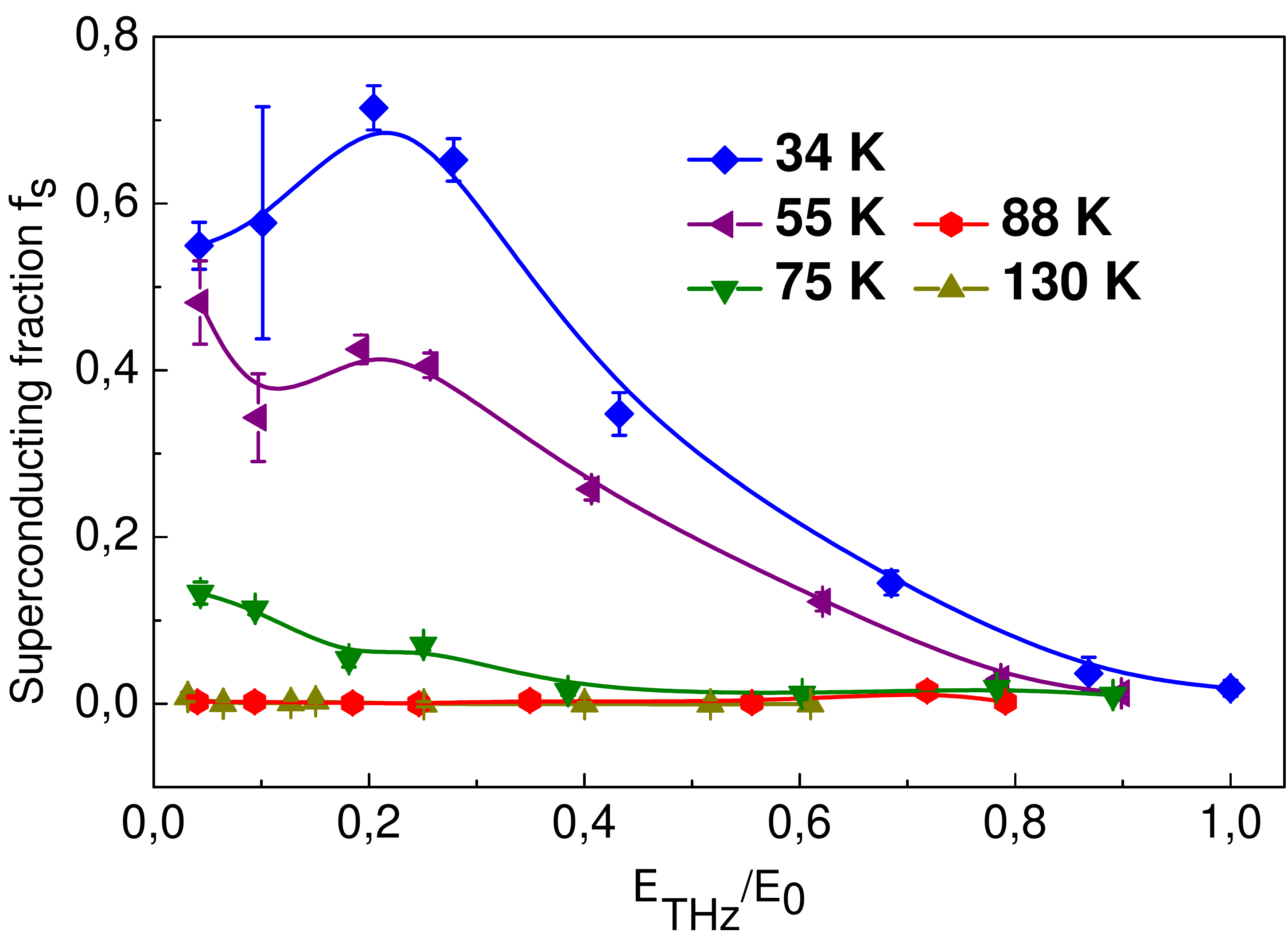}
 \caption{Superconducting fraction of the spectral weight $f_s$, extracted from fits of the two-fluid model to the complex conductivity of a YBa$_2$Cu$_3$O$_{7-\delta}$ thin film (Eq.~\ref{eq1}). Note that $f_s+f_n=1$, where $f_n$ is the quasiparticle fraction of the spectral weight. At 34\,K, 55\,K and 75\,K, the field-induced Cooper pair breakup implies a strong reduction of $f_s$. The $x$-axis is normalized to the peak THz electric field at 34\,K. The solid lines are guides to the eye.}\label{fig5}
\end{figure}

Such a behavior explains the drop of $\sigma_2$ and the increase of $\sigma_1$ for strong terahertz electric fields, as these effects clearly suggest a field-induced damping of the superconducting characteristics. It also provides an explanation for the mentioned change of the kinetic inductance, leading to an increase of the transmission for stronger electric fields (Fig.~\ref{fig3}).

Figure~\ref{fig5} complements this picture of a field-induced Cooper pair breakup and a related reduction of the superconducting carrier density. It shows the field-dependence of the superconducting fraction of the spectral weight $f_s$, extracted from a two-fluid model fit to the complex conductivity~\cite{4, 5, 9},\looseness=-1
\begin{equation}
\label{eq1}
\sigma=\sigma_1+i \sigma_2=\frac{n e^2}{m^{*}}\left( \frac{f_n}{\tau^{-1}-i\omega}-\frac{f_s}{i\omega}\right)
\end{equation}
where $n$ is the total carrier concentration, $m^{*}$ is the carriers' effective mass and $\tau$ is the scattering rate. At temperatures below $T_c$ and for increasing terahertz field strengths, $f_s$ drops sharply until the thin film becomes almost non-superconducting. This decrease of $f_s$ is accompanied by a commensurate increase of the quasiparticle contribution $f_n$.

In conclusion, we have shown how intense single-cycle terahertz pulses modify superconductivity in YBCO. A drop of the imaginary part of the conductivity and an increase of the real part imply a breakup of superconducting pairs in the presence of strong terahertz fields of tens of kV\,cm$^{-1}$. This takes place in spite of low pulse energies and photon energies well below the superconductor's energy gap. The pair breakup mechanism that explains the modification of the effective coupling function requires further investigation. The effect could be observed in all samples under investigation. In the future, it would be interesting to examine this high-field behavior in other high-temperature superconductors with a different anisotropy and to investigate a possible field-induced change of the interlayer coupling in YBCO by applying terahertz-frequency transients parallel to the $c$-axis.

\begin{acknowledgments}
We are grateful to J. Kono (Rice U.), M. Nagai (Osaka U.), S. Tani (Kyoto U.), K. Tanaka (Kyoto U.), I. Katayama (Yokohama National U.) and M. Collodo (FAU Erlangen-N\"{u}rnberg) for fruitful discussions.
\end{acknowledgments}


\end{document}